\documentclass[10pt]{article}
\usepackage{graphicx}
\usepackage{amsmath}
\usepackage{amssymb}
\usepackage{caption2}
\begin{document}
\bibliographystyle{prsty}
\begin{center}
{\large {\bf \sc{  Analysis of the tensor-tensor type scalar tetraquark states with  QCD sum rules }}} \\[2mm]
Zhi-Gang  Wang \footnote{E-mail: zgwang@aliyun.com.  }, Jun-Xia Zhang     \\
 Department of Physics, North China Electric Power University, Baoding 071003, P. R. China
\end{center}

\begin{abstract}
In this article, we study the ground states and the first radial excited states of the   tensor-tensor type scalar hidden-charm tetraquark states with the QCD sum rules. We separate the ground state contributions from the first radial excited state contributions unambiguously, and obtain the QCD sum rules for the ground states and the first radial excited states, respectively.
Then we  search for the  Borel parameters and continuum threshold
parameters    according to  four criteria and  obtain the masses  of the tensor-tensor type scalar hidden-charm tetraquark states, which  can be confronted to the experimental data in the future.
\end{abstract}

 PACS number: 12.39.Mk, 12.38.Lg

Key words: Tetraquark  states, QCD sum rules

\section{Introduction}

The attractive interaction induced by one-gluon exchange favors  formation of  diquark states in  color
 antitriplet and  disfavors formation of  diquark states in color sextet.
 The antitriplet diquark states $\varepsilon^{ijk} q^T_j C\Gamma q_k$ have  five Dirac tensor structures, scalar $C\gamma_5$,
pseudoscalar $C$, vector $C\gamma_\mu \gamma_5$, axialvector
$C\gamma_\mu $  and  tensor $C\sigma_{\mu\nu}$. The structures
$C\gamma_\mu $ and $C\sigma_{\mu\nu}$ are symmetric, while the structures
$C\gamma_5$, $C$ and $C\gamma_\mu \gamma_5$ are antisymmetric. The scalar and axialvector light diquark states have been studied with the QCD sum rules \cite{Old-Diqaurk,ZhangAL,WangLDiquark}, the scalar and axialvector heavy-light diquark states have also been studied with the QCD sum rules \cite{WangHDiquark}. The calculations based on the  QCD sum rules indicate that the scalar and axialvector diquark states are more stable than the corresponding pseudoscalar  and vector diquark states, respectively.  We usually construct the $C\gamma_5\otimes \gamma_5C$-type and $C\gamma_\mu\otimes\gamma^\mu C$-type currents to study the lowest scalar light  tetraquark states, hidden-charm or hidden-bottom  tetraquark states \cite{Nielsen-qqqq,Wang-QQ-tetra,WangTetraquarkCTP}, the corresponding $C \otimes  C$-type and $C\gamma_\mu\gamma_5\otimes\gamma_5\gamma^\mu C$-type scalar tetraquark states have much larger masses. The $C\sigma_{\alpha\beta} \otimes  \sigma^{\alpha\beta}C$-type
scalar hidden-charm or hidden-bottom tetraquark states have not been studied with the QCD sum rules, so it is interesting to study them with the QCD sum rules.

The instantons play an important role in understanding the $U_A(1)$ anomaly and
 in generating the spectrum of light hadrons \cite{Instantons}. The calculations based on the random
instanton liquid model indicate that the most strongly correlated diquarks exist in  the scalar and tensor channels \cite{RILM}. The heavy-light tensor diquark states, although they differ from the light tensor diquark states  due to the appearance of the heavy quarks, maybe play an important role in understanding the rich exotic hadron states, we should explore this  possibility, the lowest  hidden-charm and hidden-bottom tetraquark states maybe of the $C\gamma_5\otimes \gamma_5C$-type, $C\gamma_\mu\otimes\gamma^\mu C$-type or  $C\sigma_{\alpha\beta} \otimes  \sigma^{\alpha\beta}C$-type.

The  QCD sum rules provides  a powerful theoretical tool  in
 studying the hadronic properties, and has been applied extensively
  to study the masses, decay constants, hadronic form-factors,  coupling constants, etc  \cite{SVZ79,PRT85}. In this article, we construct the $C\sigma_{\alpha\beta} \otimes  \sigma^{\alpha\beta}C$-type currents to study the scalar hidden-charm tetraquark states. There exist some candidates for the scalar hidden-charm tetraquark states.
  In Ref.\cite{Lebed-3915},  Lebed and Polosa  propose that the  $X(3915)$ is the ground state scalar
$cs\bar c \bar s$ state based on  lacking of the observed $D\bar D$ and $D^*\bar{D}^*$
decays, and attribute  the single known decay mode $J/\psi \omega$ to the $\omega-\phi$ mixing effect.
Recently, the LHCb collaboration observed  two new particles $X(4500)$ and $X(4700)$ in the $J/\psi \phi$ mass spectrum  with statistical significances $6.1\sigma$ and $5.6\sigma$, respectively,  and determined the    quantum numbers  to be $J^{PC} =0^{++}$ with statistical significances  $4.0\sigma$ and $4.5\sigma$, respectively \cite{LHCb-4500-1606}.
The $X(4500)$ and $X(4700)$ are excellent  candidates for the $cs\bar{c}\bar{s}$ tetraquark states.
In Refs.\cite{Wang1606,Wang1607}, we study the $C\gamma_\mu\otimes\gamma^\mu C$-type, $C\gamma_\mu\gamma_5\otimes\gamma_5\gamma^\mu C$-type,   $C\gamma_5\otimes \gamma_5C$-type, and $C\otimes C$-type scalar $cs\bar{c}\bar{s}$ tetraquark states with the QCD sum rules. The numerical results support assigning the $X(3915)$ to be the 1S $C\gamma_5\otimes \gamma_5C$-type or $C\gamma_\mu\otimes\gamma^\mu C$-type $cs\bar{c}\bar{s}$ tetraquark state, assigning the $X(4500)$ to be the 2S  $C\gamma_\mu\otimes\gamma^\mu C$-type $cs\bar{c}\bar{s}$ tetraquark state, assigning the $X(4700)$ to be the 1S  $C\gamma_\mu\gamma_5\otimes\gamma_5\gamma^\mu C$-type $cs\bar{c}\bar{s}$ tetraquark state. For other possible assignments of the $X(4500)$ and $X(4700)$, one can consult Ref.\cite{X4500-4700}. In this article, we study the $C\sigma_{\alpha\beta} \otimes  \sigma^{\alpha\beta}C$-type hidden-charm tetraquark states with the QCD sum rules, and explore whether or not  the $X(3915)$, $X(4500)$ and $X(4700)$ can be assigned to be the  $C\sigma_{\alpha\beta} \otimes  \sigma^{\alpha\beta}C$-type tetraquark states.

The article is arranged as follows:  we derive the QCD sum rules for the masses and pole residues of  the ground state  $C\sigma_{\alpha\beta} \otimes  \sigma^{\alpha\beta}C$-type  tetraquark states  in section 2; in section 3, we derive the QCD sum rules for the masses and pole residues of  the ground state  and the first radial excited state of the $C\sigma_{\alpha\beta} \otimes  \sigma^{\alpha\beta}C$-type tetraquark states; section 4 is reserved for our conclusion.

\section{QCD sum rules for  the $C\sigma_{\alpha\beta} \otimes  \sigma^{\alpha\beta}C$-type tetraquark states without including the first radial excited states }
In the following, we write down  the two-point correlation functions $\Pi_{\bar{s}s/\bar{d}u}(p)$ in the QCD sum rules,
\begin{eqnarray}
\Pi_{\bar{s}s/\bar{d}u }(p)&=&i\int d^4x e^{ip \cdot x} \langle0|T\left\{J_{\bar{s}s/\bar{d}u}(x)J_{\bar{s}s/\bar{d}u}^{\dagger}(0)\right\}|0\rangle \, ,
\end{eqnarray}
where
\begin{eqnarray}
   J_{\bar{s}s}(x)&=&\varepsilon^{ijk}\varepsilon^{imn}s^T_j(x)C\sigma_{\alpha\beta} c_k(x) \bar{s}_m(x)\sigma^{\alpha\beta} C \bar{c}^T_n(x) \, , \nonumber\\
   J_{\bar{d}u}(x)&=&\varepsilon^{ijk}\varepsilon^{imn}u^T_j(x)C\sigma_{\alpha\beta} c_k(x) \bar{d}_m(x)\sigma^{\alpha\beta} C \bar{c}^T_n(x)\, ,
\end{eqnarray}
where the $i$, $j$, $k$, $m$, $n$ are color indexes, the $C$ is the charge conjugation matrix.

At the hadronic   side, we insert  a complete set of intermediate hadronic states with
the same quantum numbers as the current operators  $J_{\bar{s}s/\bar{d}u}(x)$ into the
correlation functions  $\Pi_{\bar{s}s/\bar{d}u}(p)$ to obtain the hadronic representation
\cite{SVZ79,PRT85}. After isolating the ground state
contributions of the scalar $cs\bar{c}\bar{s}$ tetraquark states $X_{\bar{s}s/\bar{d}u}$, we get the results,
\begin{eqnarray}
\Pi_{\bar{s}s/\bar{d}u}(p)&=&\frac{\lambda_{ \bar{s}s/\bar{d}u}^2}{M_{\bar{s}s/\bar{d}u}^2-p^2} +\cdots \, \, ,
\end{eqnarray}
where the   pole residues  $\lambda_{\bar{s}s/\bar{d}u}$ are defined by
$\langle 0|J_{\bar{s}s/\bar{d}u}(0)|X_{\bar{s}s/\bar{d}u}(p)\rangle = \lambda_{\bar{s}s/\bar{d}u}$.

 In the following,  we briefly outline  the operator product expansion for the correlation functions $\Pi_{\bar{s}s/\bar{d}u}(p)$ in perturbative QCD.  We contract the $u$, $d$, $s$  and $c$ quark fields in the correlation functions $\Pi_{\bar{s}s/\bar{d}u}(p)$ with Wick theorem, and obtain the results:
 \begin{eqnarray}
 \Pi_{\bar{s}s}(p)&=&i\varepsilon^{ijk}\varepsilon^{imn}\varepsilon^{i^{\prime}j^{\prime}k^{\prime}}\varepsilon^{i^{\prime}m^{\prime}n^{\prime}}\int d^4x e^{ip \cdot x}   \nonumber\\
&&{\rm Tr}\left[ \sigma_{\mu\nu}C^{kk^{\prime}}(x)\sigma_{\alpha\beta} CS^{jj^{\prime}T}(x)C\right] {\rm Tr}\left[ \sigma^{\alpha\beta} C^{n^{\prime}n}(-x)\sigma^{\mu\nu} C S^{m^{\prime}mT}(-x)C\right]   \, , \nonumber\\
\Pi_{\bar{d}u}(p)&=&i\varepsilon^{ijk}\varepsilon^{imn}\varepsilon^{i^{\prime}j^{\prime}k^{\prime}}\varepsilon^{i^{\prime}m^{\prime}n^{\prime}}\int d^4x e^{ip \cdot x}   \nonumber\\
&&{\rm Tr}\left[ \sigma_{\mu\nu}C^{kk^{\prime}}(x)\sigma_{\alpha\beta} CU^{jj^{\prime}T}(x)C\right] {\rm Tr}\left[ \sigma^{\alpha\beta} C^{n^{\prime}n}(-x)\sigma^{\mu\nu} C D^{m^{\prime}mT}(-x)C\right]    \, ,
\end{eqnarray}
 where the $S_{ij}(x)$, $U_{ij}(x)$, $D_{ij}(x)$ and $C_{ij}(x)$ are the full  $s$, $u$, $d$ and $c$ quark propagators, respectively,
\begin{eqnarray}
S_{ij}(x)&=& \frac{i\delta_{ij}\!\not\!{x}}{ 2\pi^2x^4}
-\frac{\delta_{ij}m_s}{4\pi^2x^2}-\frac{\delta_{ij}\langle
\bar{s}s\rangle}{12} +\frac{i\delta_{ij}\!\not\!{x}m_s
\langle\bar{s}s\rangle}{48}-\frac{\delta_{ij}x^2\langle \bar{s}g_s\sigma Gs\rangle}{192}+\frac{i\delta_{ij}x^2\!\not\!{x} m_s\langle \bar{s}g_s\sigma
 Gs\rangle }{1152}\nonumber\\
&& -\frac{ig_s G^{a}_{\alpha\beta}t^a_{ij}(\!\not\!{x}
\sigma^{\alpha\beta}+\sigma^{\alpha\beta} \!\not\!{x})}{32\pi^2x^2} -\frac{i\delta_{ij}x^2\!\not\!{x}g_s^2\langle \bar{s} s\rangle^2}{7776} -\frac{\delta_{ij}x^4\langle \bar{s}s \rangle\langle g_s^2 GG\rangle}{27648}-\frac{1}{8}\langle\bar{s}_j\sigma^{\mu\nu}s_i \rangle \sigma_{\mu\nu} \nonumber\\
&&   -\frac{1}{4}\langle\bar{s}_j\gamma^{\mu}s_i\rangle \gamma_{\mu }+\cdots \, ,  \\
{U/D}_{ij}(x)&=&S_{ij}(x)\mid_{m_s \to 0,\,\langle\bar{s}s\rangle \to \langle\bar{q}q\rangle,\,\langle\bar{s}g_s\sigma Gs\rangle \to \langle\bar{q}g_s\sigma Gq\rangle, \, \cdots}\, ,
\end{eqnarray}
\begin{eqnarray}
C_{ij}(x)&=&\frac{i}{(2\pi)^4}\int d^4k e^{-ik \cdot x} \left\{
\frac{\delta_{ij}}{\!\not\!{k}-m_c}
-\frac{g_sG^n_{\alpha\beta}t^n_{ij}}{4}\frac{\sigma^{\alpha\beta}(\!\not\!{k}+m_c)+(\!\not\!{k}+m_c)
\sigma^{\alpha\beta}}{(k^2-m_c^2)^2}\right.\nonumber\\
&&\left. +\frac{g_s D_\alpha G^n_{\beta\lambda}t^n_{ij}(f^{\lambda\beta\alpha}+f^{\lambda\alpha\beta}) }{3(k^2-m_c^2)^4}
-\frac{g_s^2 (t^at^b)_{ij} G^a_{\alpha\beta}G^b_{\mu\nu}(f^{\alpha\beta\mu\nu}+f^{\alpha\mu\beta\nu}+f^{\alpha\mu\nu\beta}) }{4(k^2-m_c^2)^5}+\cdots\right\} \, , \nonumber \\
\end{eqnarray}
\begin{eqnarray}
f^{\lambda\alpha\beta}&=&(\!\not\!{k}+m_c)\gamma^\lambda(\!\not\!{k}+m_c)\gamma^\alpha(\!\not\!{k}+m_c)\gamma^\beta(\!\not\!{k}+m_c)\, ,\nonumber\\
f^{\alpha\beta\mu\nu}&=&(\!\not\!{k}+m_c)\gamma^\alpha(\!\not\!{k}+m_c)\gamma^\beta(\!\not\!{k}+m_c)\gamma^\mu(\!\not\!{k}+m_c)\gamma^\nu(\!\not\!{k}+m_c)\, ,
\end{eqnarray}
and  $t^n=\frac{\lambda^n}{2}$, the $\lambda^n$ is the Gell-Mann matrix,  $D_\alpha=\partial_\alpha-ig_sG^n_\alpha t^n$ \cite{PRT85}.
Then we compute  the integrals both in the coordinate space and the momentum space,  and obtain the correlation functions $\Pi_{\bar{s}s/\bar{d}u}(p)$ at the quark level, therefore the QCD spectral densities through dispersion relation. In this article, we calculate the contributions of the vacuum condensates up to dimension 10 in a consistent way, for technical details, one can consult Ref.\cite{WangHuangTao}.

 Once the analytical QCD spectral densities are obtained,  we  take the
quark-hadron duality below the continuum thresholds  $s^0_{\bar{s}s/\bar{d}u}$ and perform Borel transform  with respect to
the variable $P^2=-p^2$ to obtain  the QCD sum rules:
\begin{eqnarray}
\lambda^2_{\bar{s}s/\bar{d}u}\, \exp\left(-\frac{M^2_{\bar{s}s/\bar{d}u}}{T^2}\right)= \int_{4m_c^2}^{s_{\bar{s}s/\bar{d}u}^0} ds\, \rho_{\bar{s}s/\bar{d}u}(s) \, \exp\left(-\frac{s}{T^2}\right) \, ,
\end{eqnarray}
where
\begin{eqnarray}
\rho_{\bar{s}s}(s)&=&\rho_{0}(s)+\rho_{3}(s) +\rho_{4}(s)+\rho_{5}(s)+\rho_{6}(s)+\rho_{7}(s) +\rho_{8}(s)+\rho_{10}(s)\, , \nonumber\\
\rho_{\bar{d}u}(s)&=&\rho_{\bar{s}s}(s)\mid_{m_s \to 0,\,\langle\bar{s}s\rangle \to \langle\bar{q}q\rangle,\,\langle\bar{s}g_s\sigma Gs\rangle \to \langle\bar{q}g_s\sigma Gq\rangle} \, ,
\end{eqnarray}

\begin{eqnarray}
\rho_{0}(s)&=&\frac{1}{64\pi^6}\int_{y_i}^{y_f}dy \int_{z_i}^{1-y}dz \, yz\, (1-y-z)^3\left(s-\overline{m}_c^2\right)^2\left(7s^2-6s\overline{m}_c^2+\overline{m}_c^4 \right)    \nonumber\\
&&+\frac{1}{32\pi^6}\int_{y_i}^{y_f}dy \int_{z_i}^{1-y}dz \, yz\, (1-y-z)^2\left(s-\overline{m}_c^2\right)^3\left(3s-\overline{m}_c^2 \right) \, ,
\end{eqnarray}

\begin{eqnarray}
\rho_{3}(s)&=&\frac{m_s\langle \bar{s}s\rangle}{2\pi^4} \int_{y_i}^{y_f}dy \int_{z_i}^{1-y}dz \, yz\, (1-y-z) \left(10s^2-12s\overline{m}_c^2+3\overline{m}_c^4 \right)   \nonumber\\
&&+\frac{m_s\langle \bar{s}s\rangle}{\pi^4}\int_{y_i}^{y_f}dy \int_{z_i}^{1-y}dz \, yz\, \left(s-\overline{m}_c^2 \right)\left(2s-\overline{m}_c^2 \right)\nonumber\\
&&-\frac{3m_s m_c^2\langle \bar{s}s\rangle}{\pi^4}\int_{y_i}^{y_f}dy \int_{z_i}^{1-y}dz  \left( s - \overline{m}_c^2\right) \, ,
\end{eqnarray}

\begin{eqnarray}
\rho_{4}(s)&=&-\frac{m_c^2}{48\pi^4} \langle\frac{\alpha_s GG}{\pi}\rangle\int_{y_i}^{y_f}dy \int_{z_i}^{1-y}dz\, \left( \frac{z}{y^2}+\frac{y}{z^2}\right)(1-y-z)^3 \left\{ 2s-\overline{m}_c^2+\frac{s^2}{6}\delta\left(s-\overline{m}_c^2\right)\right\} \nonumber\\
&&-\frac{m_c^2}{48\pi^4} \langle\frac{\alpha_s GG}{\pi}\rangle\int_{y_i}^{y_f}dy \int_{z_i}^{1-y}dz \,\left( \frac{z}{y^2}+\frac{y}{z^2}\right)(1-y-z)^2 \left( 3s-2\overline{m}_c^2\right) \nonumber\\
&&-\frac{1}{864\pi^4} \langle\frac{\alpha_s GG}{\pi}\rangle\int_{y_i}^{y_f}dy \int_{z_i}^{1-y}dz \,(1-y-z)^3 \left( 10s^2-12s\overline{m}_c^2+3\overline{m}_c^4\right)  \nonumber\\
&&+\frac{7}{288\pi^4} \langle\frac{\alpha_s GG}{\pi}\rangle\int_{y_i}^{y_f}dy \int_{z_i}^{1-y}dz  \,(1-y-z)^2 \left( s-\overline{m}_c^2\right)\left( 2s-\overline{m}_c^2\right)  \nonumber\\
&&+\frac{7}{576\pi^4} \langle\frac{\alpha_s GG}{\pi}\rangle\int_{y_i}^{y_f}dy \int_{z_i}^{1-y}dz \,\left( y+z\right)(1-y-z)^2 \left( 10s^2-12s\overline{m}_c^2+3\overline{m}_c^4\right)  \nonumber\\
&&-\frac{1}{72\pi^4} \langle\frac{\alpha_s GG}{\pi}\rangle\int_{y_i}^{y_f}dy \int_{z_i}^{1-y}dz \,\left( y+z\right)(1-y-z) \left( s-\overline{m}_c^2\right)\left( 2s-\overline{m}_c^2\right)  \nonumber\\
&&-\frac{1}{144\pi^4} \langle\frac{\alpha_s GG}{\pi}\rangle\int_{y_i}^{y_f}dy \int_{z_i}^{1-y}dz  \,yz(1-y-z) \left( 10s^2-12s\overline{m}_c^2+3\overline{m}_c^4\right)  \nonumber\\
&&+\frac{7}{144\pi^4} \langle\frac{\alpha_s GG}{\pi}\rangle\int_{y_i}^{y_f}dy \int_{z_i}^{1-y}dz \,yz  \left( s-\overline{m}_c^2\right)\left( 2s-\overline{m}_c^2\right) \, ,
\end{eqnarray}

\begin{eqnarray}
\rho_{5}(s)&=&-\frac{m_s\langle \bar{s}g_s\sigma Gs\rangle}{2\pi^4}\int_{y_i}^{y_f}dy \int_{z_i}^{1-y}dz  \,yz  \left\{2s-\overline{m}_c^2 +\frac{s^2}{6}\delta\left(s-\overline{m}_c^2 \right)\right\}    \nonumber\\
&&-\frac{m_s\langle \bar{s}g_s\sigma Gs\rangle}{6\pi^4}\int_{y_i}^{y_f}dy \, y(1-y) \left(3s-2\widetilde{m}_c^2 \right) +\frac{3m_s m_c^2\langle \bar{s}g_s\sigma Gs\rangle}{4\pi^4}\int_{y_i}^{y_f}dy     \nonumber\\
&&-\frac{m_s m_c^2\langle \bar{s}g_s\sigma Gs\rangle}{24\pi^4}\int_{y_i}^{y_f}dy \int_{z_i}^{1-y}dz  \, \left( \frac{1}{y}+\frac{1}{z}\right)  \, ,
\end{eqnarray}

\begin{eqnarray}
\rho_{6}(s)&=&\frac{2m_c^2\langle\bar{s}s\rangle^2}{\pi^2}\int_{y_i}^{y_f}dy   +\frac{2g_s^2\langle\bar{s}s\rangle^2}{27\pi^4}\int_{y_i}^{y_f}dy \int_{z_i}^{1-y}dz\, yz \left\{2s-\overline{m}_c^2 +\frac{s^2}{6}\delta\left(s-\overline{m}_c^2 \right)\right\}\nonumber\\
&&+\frac{2g_s^2\langle\bar{s}s\rangle^2}{81\pi^4}\int_{y_i}^{y_f}dy \, y(1-y) \left(3s-2\widetilde{m}_c^2 \right)\nonumber\\
&&-\frac{g_s^2\langle\bar{s}s\rangle^2}{162\pi^4}\int_{y_i}^{y_f}dy \int_{z_i}^{1-y}dz \, (1-y-z)\left\{ \frac{10}{3}\left(\frac{z}{y^2}+\frac{y}{z^2} \right)m_c^2\left[ 2+ s\,\delta\left(s-\overline{m}_c^2 \right)\right] \right. \nonumber\\
&&\left.+28(y+z) \left[2s-\overline{m}_c^2 +\frac{s^2}{6}\delta\left(s-\overline{m}_c^2 \right)\right]+9\left(\frac{z}{y}+\frac{y}{z} \right)\left(3s-2\overline{m}_c^2 \right)\right\}  \, ,
\end{eqnarray}

\begin{eqnarray}
\rho_7(s)&=&-\frac{m_s m_c^2\langle\bar{s}s\rangle}{18\pi^2  }\langle\frac{\alpha_sGG}{\pi}\rangle\int_{0}^{1}dy \int_{0}^{1-y}dz \left(\frac{z}{y^2}+\frac{y}{z^2} \right)(1-y-z)\left(1+\frac{ s}{T^2}+\frac{ s^2}{2T^4}\right) \delta\left(s-\overline{m}_c^2\right)\nonumber\\
&&-\frac{m_s m_c^2\langle\bar{s}s\rangle}{18\pi^2  }\langle\frac{\alpha_sGG}{\pi}\rangle\int_{0}^{1}dy \int_{0}^{1-y}dz \left(\frac{z}{y^2}+\frac{y}{z^2} \right) \left(1+\frac{ s}{T^2}\right) \delta\left(s-\overline{m}_c^2\right)\nonumber\\
&&+\frac{m_s m_c^4\langle\bar{s}s\rangle}{6\pi^2T^2  }\langle\frac{\alpha_sGG}{\pi}\rangle\int_{0}^{1}dy \int_{0}^{1-y}dz \left(\frac{1}{y^3}+\frac{1}{z^3} \right) \delta\left(s-\overline{m}_c^2\right)\nonumber\\
&&-\frac{m_s m_c^2\langle\bar{s}s\rangle}{2\pi^2  }\langle\frac{\alpha_sGG}{\pi}\rangle\int_{0}^{1}dy \int_{0}^{1-y}dz \left(\frac{1}{y^2}+\frac{1}{z^2} \right) \delta\left(s-\overline{m}_c^2\right)\nonumber\\
&&-\frac{m_s \langle\bar{s}s\rangle}{18\pi^2  }\langle\frac{\alpha_sGG}{\pi}\rangle\int_{y_i}^{y_f}dy \int_{z_i}^{1-y}dz \left(1-y-z \right) \left\{ 1+\left( \frac{2s}{3}+\frac{s^2}{6T^2}\right)\delta\left(s-\overline{m}_c^2\right)\right\}\nonumber\\
&&+\frac{m_s \langle\bar{s}s\rangle}{9\pi^2  }\langle\frac{\alpha_sGG}{\pi}\rangle\int_{0}^{1}dy  \left\{ 1+\frac{s}{2}\delta\left(s-\widetilde{m}_c^2\right)\right\} \nonumber\\
&&+\frac{7m_s \langle\bar{s}s\rangle}{72\pi^2  }\langle\frac{\alpha_sGG}{\pi}\rangle\int_{y_i}^{y_f}dy \int_{z_i}^{1-y}dz \left(y+z \right) \left\{ 1+\left( \frac{2s}{3}+\frac{s^2}{6T^2}\right)\delta\left(s-\overline{m}_c^2\right)\right\}\nonumber\\
&&-\frac{2m_s m_c^2 \langle\bar{s}s\rangle}{9\pi^2  }\langle\frac{\alpha_sGG}{\pi}\rangle\int_{y_i}^{y_f}dy \int_{z_i}^{1-y}dz \,\frac{1}{yz} \, \delta\left(s-\overline{m}_c^2\right)\nonumber\\
&&-\frac{m_s m_c^2\langle\bar{s}s\rangle}{12\pi^2  }\langle\frac{\alpha_sGG}{\pi}\rangle\int_{0}^{1}dy  \left( 1+\frac{s}{T^2}\right)\delta\left(s-\widetilde{m}_c^2\right) \, ,
\end{eqnarray}

\begin{eqnarray}
\rho_8(s)&=&\frac{ \langle\bar{s}s\rangle\langle\bar{s}g_s\sigma Gs\rangle}{18\pi^2}\int_{0}^{1} dy \,s\,\delta\left(s-\widetilde{m}_c^2\right)-\frac{m_c^2\langle\bar{s}s\rangle\langle\bar{s}g_s\sigma Gs\rangle}{\pi^2}\int_0^1 dy \left(1+\frac{s}{T^2} \right)\delta\left(s-\widetilde{m}_c^2\right)\, , \nonumber\\
\end{eqnarray}

\begin{eqnarray}
\rho_{10}(s)&=&\frac{m_c^2\langle\bar{s}g_s\sigma Gs\rangle^2}{8\pi^2T^6}\int_0^1 dy \, s^2 \, \delta \left( s-\widetilde{m}_c^2\right)
\nonumber \\
&&-\frac{m_c^4\langle\bar{s}s\rangle^2}{9T^4}\langle\frac{\alpha_sGG}{\pi}\rangle\int_0^1 dy  \left\{ \frac{1}{y^3}+\frac{1}{(1-y)^3}\right\} \delta\left( s-\widetilde{m}_c^2\right)\nonumber\\
&&+\frac{m_c^2\langle\bar{s}s\rangle^2}{3T^2}\langle\frac{\alpha_sGG}{\pi}\rangle\int_0^1 dy  \left\{ \frac{1}{y^2}+\frac{1}{(1-y)^2}\right\} \delta\left( s-\widetilde{m}_c^2\right)\nonumber\\
&&+\frac{4 \langle\bar{s}s\rangle^2}{27 T^2}\langle\frac{\alpha_sGG}{\pi}\rangle \int_0^1 dy   \,s\,   \delta\left( s-\widetilde{m}_c^2\right)+\frac{11 \langle\bar{s}g_s\sigma Gs\rangle^2}{36\pi^2T^2}\int_0^1 dy \, s \, \delta \left( s-\widetilde{m}_c^2\right)\nonumber \\
&&-\frac{ \langle\bar{s}g_s\sigma Gs\rangle^2}{72\pi^2T^4}\int_0^1 dy \, s^2 \, \delta \left( s-\widetilde{m}_c^2\right)+\frac{m_c^2\langle\bar{s}s\rangle^2}{9T^6}\langle\frac{\alpha_sGG}{\pi}\rangle\int_0^1 dy  \,s^2\, \delta\left( s-\widetilde{m}_c^2\right)\, ,
\end{eqnarray}
 $y_{f}=\frac{1+\sqrt{1-4m_c^2/s}}{2}$,
$y_{i}=\frac{1-\sqrt{1-4m_c^2/s}}{2}$, $z_{i}=\frac{y
m_c^2}{y s -m_c^2}$, $\overline{m}_c^2=\frac{(y+z)m_c^2}{yz}$,
$ \widetilde{m}_c^2=\frac{m_c^2}{y(1-y)}$, $\int_{y_i}^{y_f}dy \to \int_{0}^{1}dy$, $\int_{z_i}^{1-y}dz \to \int_{0}^{1-y}dz$, when the $\delta$ functions $\delta\left(s-\overline{m}_c^2\right)$ and $\delta\left(s-\widetilde{m}_c^2\right)$ appear.

 We differentiate   Eq.(9) with respect to  $\frac{1}{T^2}$, then eliminate the
 pole residues $\lambda_{\bar{s}s/\bar{d}u}$, and  obtain the QCD sum rules for
 the ground state masses $ M_{\bar{s}s/\bar{d}u}$ of the  $C\sigma_{\alpha\beta} \otimes  \sigma^{\alpha\beta}C$-type scalar  hidden-charm tetraquark states,
 \begin{eqnarray}
 M^2_{\bar{s}s/\bar{d}u}=- \frac{\int_{4m_c^2}^{s^0_{\bar{s}s/\bar{d}u}} ds\frac{d}{d \left(1/T^2\right)}\rho_{\bar{s}s/\bar{d}u}(s)\exp\left(-\frac{s}{T^2}\right)}{\int_{4m_c^2}^{s_{\bar{s}s/\bar{d}u}^0} ds \rho_{\bar{s}s/\bar{d}u}(s)\exp\left(-\frac{s}{T^2}\right)}\, .
\end{eqnarray}

The input parameters are shown explicitly in Table 1.
The quark condensates, mixed quark condensates and $\overline{MS}$ masses  evolve with the   renormalization group equation, we take into account
the energy-scale dependence according to the following equations,
\begin{eqnarray}
\langle\bar{q}q \rangle(\mu)&=&\langle\bar{q}q \rangle(Q)\left[\frac{\alpha_{s}(Q)}{\alpha_{s}(\mu)}\right]^{\frac{4}{9}}\, , \nonumber\\
 \langle\bar{s}s \rangle(\mu)&=&\langle\bar{s}s \rangle(Q)\left[\frac{\alpha_{s}(Q)}{\alpha_{s}(\mu)}\right]^{\frac{4}{9}}\, , \nonumber\\
 \langle\bar{q}g_s \sigma Gq \rangle(\mu)&=&\langle\bar{q}g_s \sigma Gq \rangle(Q)\left[\frac{\alpha_{s}(Q)}{\alpha_{s}(\mu)}\right]^{\frac{2}{27}}\, , \nonumber\\ \langle\bar{s}g_s \sigma Gs \rangle(\mu)&=&\langle\bar{s}g_s \sigma Gs \rangle(Q)\left[\frac{\alpha_{s}(Q)}{\alpha_{s}(\mu)}\right]^{\frac{2}{27}}\, , \nonumber\\
m_c(\mu)&=&m_c(m_c)\left[\frac{\alpha_{s}(\mu)}{\alpha_{s}(m_c)}\right]^{\frac{12}{25}} \, ,\nonumber\\
m_s(\mu)&=&m_s({\rm 2GeV} )\left[\frac{\alpha_{s}(\mu)}{\alpha_{s}({\rm 2GeV})}\right]^{\frac{4}{9}} \, ,\nonumber\\
\alpha_s(\mu)&=&\frac{1}{b_0t}\left[1-\frac{b_1}{b_0^2}\frac{\log t}{t} +\frac{b_1^2(\log^2{t}-\log{t}-1)+b_0b_2}{b_0^4t^2}\right]\, ,
\end{eqnarray}
  where $t=\log \frac{\mu^2}{\Lambda^2}$, $b_0=\frac{33-2n_f}{12\pi}$, $b_1=\frac{153-19n_f}{24\pi^2}$, $b_2=\frac{2857-\frac{5033}{9}n_f+\frac{325}{27}n_f^2}{128\pi^3}$,  $\Lambda=213\,\rm{MeV}$, $296\,\rm{MeV}$  and  $339\,\rm{MeV}$ for the flavors  $n_f=5$, $4$ and $3$, respectively  \cite{PDG}.
 Furthermore, we set  $m_u=m_d=0$.

\begin{table}
\begin{center}
\begin{tabular}{|c|c|c|c|}\hline\hline
    Parameters                                          & Values\\   \hline
   $\langle\bar{q}q \rangle({\rm 1GeV})$                & $-(0.24\pm 0.01\, \rm{GeV})^3$ \,\, \cite{SVZ79,PRT85,ColangeloReview}         \\  \hline
   $\langle\bar{s}s \rangle({\rm 1GeV})$                & $(0.8\pm0.1)\langle\bar{q}q \rangle({\rm 1GeV})$ \,\, \cite{SVZ79,PRT85,ColangeloReview}     \\ \hline
$\langle\bar{q}g_s\sigma G q \rangle({\rm 1GeV})$       & $m_0^2\langle \bar{q}q \rangle({\rm 1GeV})$   \,\,  \cite{SVZ79,PRT85,ColangeloReview}       \\  \hline
$\langle\bar{s}g_s\sigma G s \rangle({\rm 1GeV})$       & $m_0^2\langle \bar{s}s \rangle({\rm 1GeV})$  \,\,  \cite{SVZ79,PRT85,ColangeloReview}        \\  \hline
$m_0^2({\rm 1GeV})$                                     & $(0.8 \pm 0.1)\,\rm{GeV}^2$      \,\,  \cite{SVZ79,PRT85,ColangeloReview}    \\   \hline
 $\langle \frac{\alpha_s GG}{\pi}\rangle$               & $(0.33\,\rm{GeV})^4$          \,\,  \cite{SVZ79,PRT85,ColangeloReview} \\   \hline
   $m_{c}(m_c)$                                         & $(1.275\pm0.025)\,\rm{GeV}$ \,\, \cite{PDG}      \\    \hline
   $m_{s}({\rm 2GeV})$                                  & $(0.095\pm0.005)\,\rm{GeV}$  \,\, \cite{PDG}      \\   \hline \hline
\end{tabular}
\end{center}
\caption{ The  input parameters in the QCD sum rules, the values in the bracket denote the energy scales $\mu=1\,\rm{GeV}$, $2\,\rm{GeV}$ and $m_c$, respectively. }
\end{table}

  In the diquark-antidiquark type tetraquark system $Qq\bar{Q}\bar{q}^{\prime}$,
 the $Q$-quark serves as a static well potential and  combines with the light quark $q$  to form a heavy diquark $\mathcal{D}$ in  color antitriplet,
while the $\bar{Q}$-quark serves  as another static well potential and combines with the light antiquark $\bar{q}^\prime$  to form a heavy antidiquark $\mathcal{\bar{D}}$ in  color triplet;  the  $\mathcal{D}$ and $\mathcal{\bar{D}}$ combine together to form a compact tetraquark state \cite{WangTetraquarkCTP,WangHuangTao,Wang-4660-2014,Wang-Huang-NPA-2014}. For such diquark-antidiquark type tetraquark systems, we suggest an   energy  scale formula $\mu=\sqrt{M^2_{X/Y/Z}-(2{\mathbb{M}}_Q)^2}$ to determine the energy scales of the QCD spectral densities, where the $X$, $Y$ and $Z$ are the hidden-charm or hidden-bottom tetraquark states $Qq\bar{Q}\bar{q}^{\prime}$, the ${\mathbb{M}}_Q$ are the effective heavy quark masses. In this article, we choose  the updated value  ${\mathbb{M}}_c=1.82\,\rm{GeV}$ \cite{WangEPJC1601}.

 Now  we search for the  Borel parameters $T^2$ and continuum threshold
parameters $s^0_{\bar{s}s/\bar{d}u}$  according to  the  four criteria:

$\bf{1_\cdot}$ Pole dominance at the hadron  side;

$\bf{2_\cdot}$ Convergence of the operator product expansion;

$\bf{3_\cdot}$ Appearance of the Borel platforms;

$\bf{4_\cdot}$ Satisfying the energy scale formula.\\
We cannot obtain reasonable Borel parameters $T^2$ and continuum threshold parameters $s^0_{\bar{s}s/\bar{d}u}$,  if the energy gap between the ground state and the first radial excited state is about $0.3-0.7\,\rm{GeV}$.

\section{QCD sum rules for  the $C\sigma_{\alpha\beta} \otimes  \sigma^{\alpha\beta}C$-type tetraquark states  including the first radial excited states }

Now we take into account both  the ground state contribution and the first radial excited state contribution at the hadronic side of the QCD sum rules \cite{Baxi-G}.
Firstly,  we   introduce the notations $\tau=\frac{1}{T^2}$, $D^n=\left( -\frac{d}{d\tau}\right)^n$, and use the subscripts $1$ and $2$ to denote the ground state and the first radial excited state of the $C\sigma_{\alpha\beta} \otimes  \sigma^{\alpha\beta}C$-type tetraquark states, respectively.
 Then the QCD sum rules    can be written as
\begin{eqnarray}
\lambda_1^2\exp\left(-\tau M_1^2 \right)+\lambda_2^2\exp\left(-\tau M_2^2 \right)&=&\Pi_{QCD}(\tau) \, ,
\end{eqnarray}
 the subscript $QCD$  denotes the QCD side of the correlation functions $\Pi_{\bar{s}s/\bar{d}u}(\tau)$.
We differentiate both sides of the QCD sum rules in Eq.(21) with respect to $\tau$ and obtain
\begin{eqnarray}
\lambda_1^2M_1^2\exp\left(-\tau M_1^2 \right)+\lambda_2^2M_2^2\exp\left(-\tau M_2^2 \right)&=&D\Pi_{QCD}(\tau) \, .
\end{eqnarray}
Then we solve the two equations, and obtain the QCD sum rules,
\begin{eqnarray}
\lambda_i^2\exp\left(-\tau M_i^2 \right)&=&\frac{\left(D-M_j^2\right)\Pi_{QCD}(\tau)}{M_i^2-M_j^2} \, ,
\end{eqnarray}
where $i \neq j$.
We differentiate both sides of the QCD sum rules in Eq.(23) with respect to $\tau$ and obtain
\begin{eqnarray}
M_i^2&=&\frac{\left(D^2-M_j^2D\right)\Pi_{QCD}(\tau)}{\left(D-M_j^2\right)\Pi_{QCD}(\tau)} \, , \nonumber\\
M_i^4&=&\frac{\left(D^3-M_j^2D^2\right)\Pi_{QCD}(\tau)}{\left(D-M_j^2\right)\Pi_{QCD}(\tau)}\, .
\end{eqnarray}
 The squared masses $M_i^2$ satisfy the following equation,
\begin{eqnarray}
M_i^4-b M_i^2+c&=&0\, ,
\end{eqnarray}
where
\begin{eqnarray}
b&=&\frac{D^3\otimes D^0-D^2\otimes D}{D^2\otimes D^0-D\otimes D}\, , \nonumber\\
c&=&\frac{D^3\otimes D-D^2\otimes D^2}{D^2\otimes D^0-D\otimes D}\, , \nonumber\\
D^j \otimes D^k&=&D^j\Pi_{QCD}(\tau) \,  D^k\Pi_{QCD}(\tau)\, ,
\end{eqnarray}
$i=1,2$, $j,k=0,1,2,3$.
We solve the equation in Eq.(25) and obtain two solutions
\begin{eqnarray}
M_1^2=\frac{b-\sqrt{b^2-4c} }{2} \, , \\
M_2^2=\frac{b+\sqrt{b^2-4c} }{2} \, .
\end{eqnarray}
 The ground state contributions are  separated from the first radial excited state contributions unambiguously, and we obtain the QCD sum rules for the ground states and the first radial excited states, respectively.

Again, we search for the  Borel parameters $T^2$ and continuum threshold
parameters $s^0_{\bar{s}s/\bar{d}u}$  according to  the  four criteria:

$\bf{1_\cdot}$ Pole dominance at the hadron  side;

$\bf{2_\cdot}$ Convergence of the operator product expansion;

$\bf{3_\cdot}$ Appearance of the Borel platforms;

$\bf{4_\cdot}$ Satisfying the energy scale formula.\\
The resulting Borel parameters $T^2$ and continuum threshold
parameters $s^0_{\bar{s}s/\bar{d}u}$ are
\begin{eqnarray}
  X_{\bar{d}u} &:& T^2 = (2.0-2.4) \mbox{ GeV}^2 \, ,\, s^0_{\bar{d}u} = (4.8\pm0.1 \mbox{ GeV})^2 \, , \nonumber\\
  X_{\bar{s}s} &:& T^2 = (2.1-2.5) \mbox{ GeV}^2 \, ,\, s^0_{\bar{s}s} = (4.8\pm0.1 \mbox{ GeV})^2 \, .
  \end{eqnarray}
The pole contributions are
\begin{eqnarray}
   X_{\bar{d}u}({\rm 1S+2S} ) &:& {\rm pole}  = (68-89) \% \, \,\, {\rm at}\, \,\, \mu = 1.20 \mbox{ GeV} \, , \nonumber\\
   X_{\bar{d}u}({\rm 1S+2S} ) &:& {\rm pole}  = (79-95) \% \, \,\, {\rm at}\, \,\, \mu = 2.45 \mbox{ GeV} \, , \\
   X_{\bar{s}s}({\rm 1S+2S} ) &:& {\rm pole}  = (64-87) \% \, \,\, {\rm at}\, \,\, \mu = 1.20 \mbox{ GeV} \, , \nonumber \\
   X_{\bar{s}s}({\rm 1S+2S} ) &:& {\rm pole}  = (76-93) \% \, \,\, {\rm at}\, \,\, \mu = 2.50 \mbox{ GeV} \, ,
\end{eqnarray}
the pole dominance condition is well satisfied, the criterion $\bf{1}$ is satisfied.
The contributions come from the vacuum condensates of dimension 10 $D_{10}$ are
 \begin{eqnarray}
   X_{\bar{d}u}({\rm 1S+2S} ) &:& D_{10}  = (9-25) \% \, \,\, {\rm at}\, \,\, \mu = 1.20 \mbox{ GeV} \, , \nonumber\\
   X_{\bar{d}u}({\rm 1S+2S} ) &:& D_{10}   = (4-11) \% \, \,\, {\rm at}\, \,\, \mu = 2.45 \mbox{ GeV} \, , \\
   X_{\bar{s}s}({\rm 1S+2S} ) &:& D_{10}   = (5-14) \% \, \,\, {\rm at}\, \,\, \mu = 1.20 \mbox{ GeV} \, , \nonumber \\
   X_{\bar{s}s}({\rm 1S+2S} ) &:& D_{10}   = (2-6) \% \, \,\, {\rm at}\, \,\, \mu = 2.50 \mbox{ GeV} \, ,
\end{eqnarray}
for the central values of the continuum threshold parameters, the operator product expansion is  convergent, the criterion $\bf{2}$ is satisfied.

 Now we take into account the uncertainties of all the input parameters, and obtain the masses and pole residues of the $C\sigma_{\alpha\beta} \otimes  \sigma^{\alpha\beta}C$-type tetraquark states,
\begin{eqnarray}
M_{\bar{d}u,{\rm 1S}}&=&3.82\pm0.16 \,\rm{GeV} \, ,  \nonumber\\
M_{\bar{s}s,{\rm 1S}}&=&3.84\pm0.16 \,\rm{GeV} \, ,  \nonumber\\
\lambda_{\bar{d}u,{\rm 1S}}&=&\left( 5.20\pm1.35\right) \times 10^{-2}\,\rm{GeV}^5 \,   ,\nonumber\\
\lambda_{\bar{s}s,{\rm 1S}}&=&\left(4.87\pm 1.25\right)\times 10^{-2}\,\rm{GeV}^5 \,   ,
\end{eqnarray}
at the energy scale $\mu=1.20\,\rm{GeV}$,
\begin{eqnarray}
M_{\bar{d}u,{\rm 2S}}&=&4.38\pm0.09 \,\rm{GeV} \, ,  \nonumber\\
\lambda_{\bar{d}u,{\rm 2S}}&=&\left(2.12\pm0.31\right)\times 10^{-1}\,\rm{GeV}^5 \, ,
\end{eqnarray}
at the energy scale $\mu=2.45\,\rm{GeV}$,
\begin{eqnarray}
M_{\bar{s}s,{\rm 2S}}&=&4.40\pm0.09 \,\rm{GeV} \, ,  \nonumber\\
\lambda_{\bar{s}s,{\rm 2S}}&=&\left(2.14\pm0.33\right)\times 10^{-1}\,\rm{GeV}^5 \,   ,
\end{eqnarray}
at the energy scale $\mu=2.50\,\rm{GeV}$. The central values of the predicted masses satisfy the energy scale formula, the criterion $\bf{4}$ is satisfied.

In Fig.1, we plot the predicted masses $M_{\bar{d}u/\bar{s}s}$ with variations of the Borel parameters $T^2$. From the figure, we can see that the plateaus are rather flat, the criterion $\bf{3}$ is satisfied. The four criteria are all satisfied, we expect to make reliable predictions.

\begin{figure}
 \centering
\includegraphics[totalheight=5cm,width=6cm]{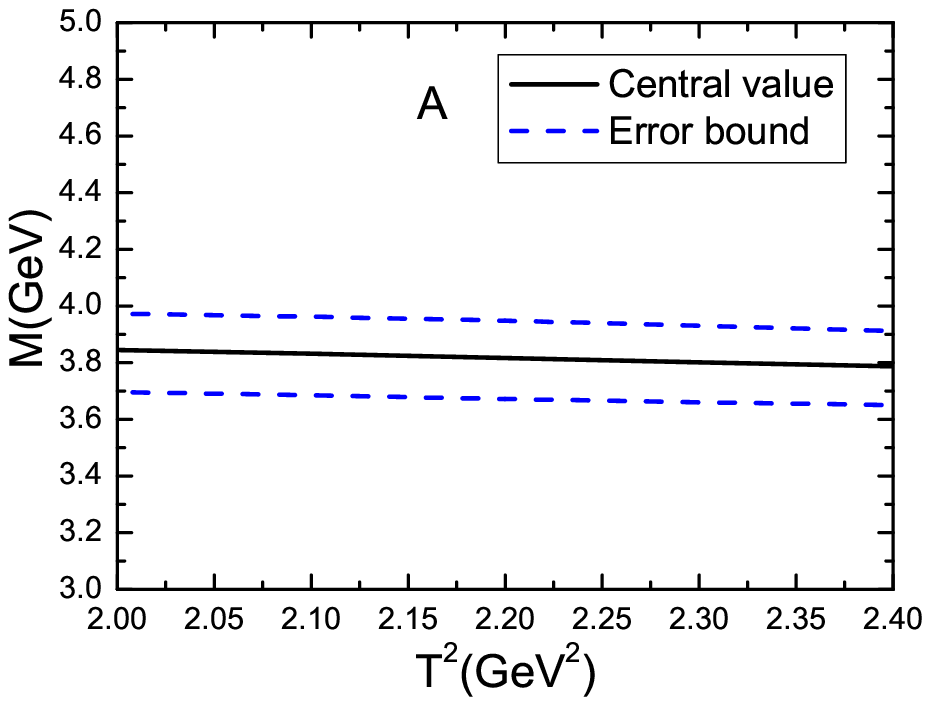}
\includegraphics[totalheight=5cm,width=6cm]{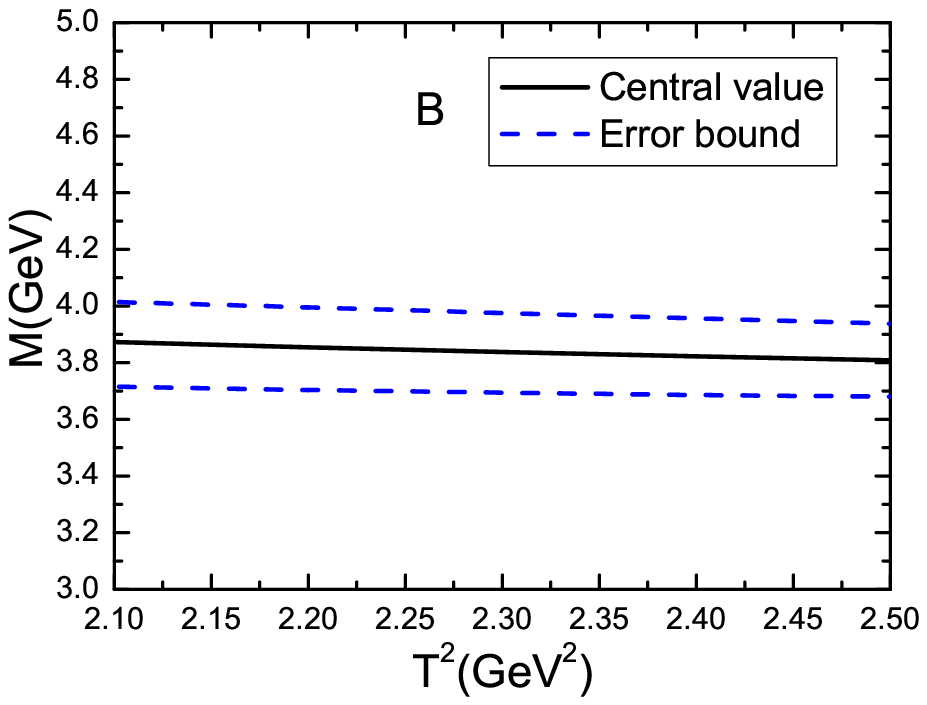}
\includegraphics[totalheight=5cm,width=6cm]{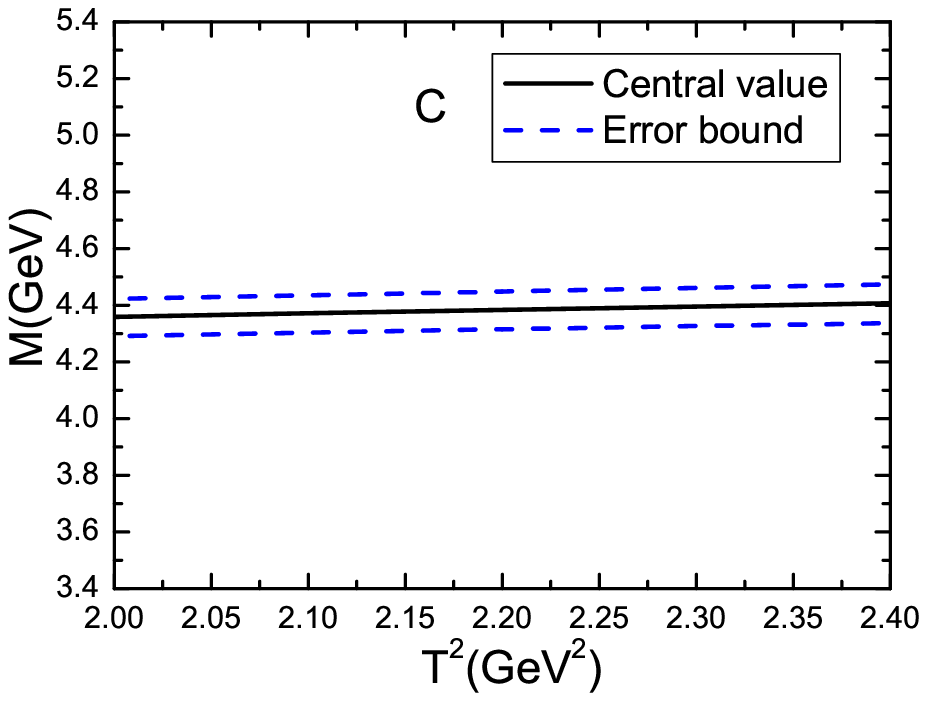}
\includegraphics[totalheight=5cm,width=6cm]{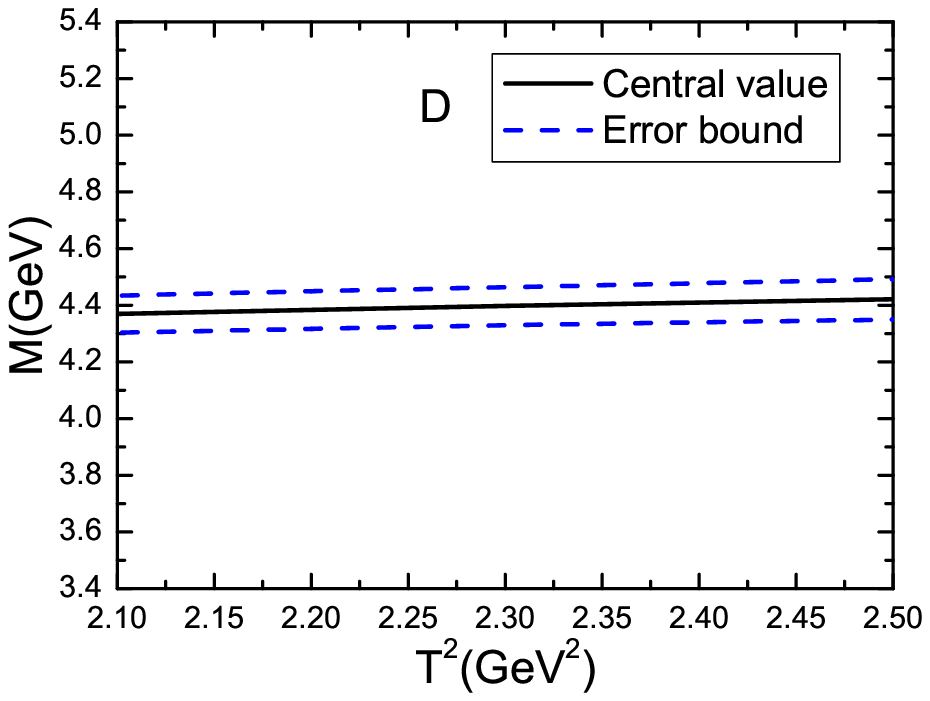}
      \caption{  The masses with variations of the Borel parameters, where the  $A$, $B$, $C$ and $D$ denote the $X_{\bar{d}u,{\rm 1S}}$, $X_{\bar{s}s,{\rm 1S}}$,
        $X_{\bar{d}u,{\rm 2S}}$ and $X_{\bar{s}s,{\rm 2S}}$, respectively.}
\end{figure}

The energy gaps between the ground states and the first radial excited states are
\begin{eqnarray}
M_{\bar{d}u,{\rm 2S}}-M_{\bar{d}u,{\rm 1S}}&=&0.56\,\rm{GeV}\, , \nonumber\\
M_{\bar{s}s,{\rm 2S}}-M_{\bar{s}s,{\rm 1S}}&=&0.56\,\rm{GeV}\, .
\end{eqnarray}
The  $Z(4430)$ is assigned to be  the first radial excitation of the $Z_c(3900)$ according to the
analogous decays,
\begin{eqnarray}
Z_c(3900)^\pm&\to&J/\psi\pi^\pm\, , \nonumber \\
Z(4430)^\pm&\to&\psi^\prime\pi^\pm\, ,
\end{eqnarray}
and the  mass differences $M_{Z(4430)}-M_{Z_c(3900)}=576\,\rm{MeV}$ and $M_{\psi^\prime}-M_{J/\psi}=589\,\rm{MeV}$  \cite{Z4430-1405,Wang4430}.
The energy gaps $M_{\bar{d}u,{\rm 2S}}-M_{\bar{d}u,{\rm 1S}}$, $M_{\bar{s}s,{\rm 2S}}-M_{\bar{s}s,{\rm 1S}}$, $M_{Z(4430)}-M_{Z_c(3900)}$ are compatible with each other.
The widths $\Gamma_{Z_c(3900)}=46\pm 10 \pm20 \, \rm {MeV}$ \cite{Zc3900} and  $\Gamma_{Z(4430)} =172 \pm 13 {}^{+37}_{-34} \,\rm{MeV}$ \cite{Z4430-LHCb} are not broad,
 the QCD sum rules for the ground state $Z_c(3900)$ alone or without including the first radial excited state $Z(4430)$ work well \cite{WangHuangTao}. If both the ground state $Z_c(3900)$ and the first radial excited state $Z(4430)$ are included in, the continuum threshold parameter $\sqrt{s_0}=4.8\pm 0.1\,{\rm{GeV}}= M_{Z(4430)}+(0.2\sim0.4) \,\rm{GeV}$, the lower bound of the $\sqrt{s_0}-M_{Z(4430)}$ is about $0.2\,\rm{GeV}$, which is large enough to take into account  the contribution of the $Z(4430)$   \cite{Wang4430}. In the present case, the lower bound of the $\sqrt{s_0}-M_{\bar{d}u/\bar{s}s,2\rm{S}}$ are about $0.3\,\rm{GeV}$, which indicates  that the widths of the first radial excited states of the $C\sigma_{\alpha\beta} \otimes  \sigma^{\alpha\beta}C$-type tetraquark states are rather large.  According to the energy gaps between the ground states and the first radial excited states,   the continuum threshold parameters should be chosen as large as $\sqrt{s_0}-M_{\bar{d}u/\bar{s}s,1\rm{S}}=0.5\pm0.1\,\rm{GeV}$ without including the first radial excited states $X_{\bar{d}u/\bar{s}s,2\rm{S}}$ explicitly, however,  for such large continuum thresholds, the contributions of the $X_{\bar{d}u/\bar{s}s,2\rm{S}}$ are already included in due to their large widths. So the QCD sum rules in which only the ground state $C\sigma_{\alpha\beta} \otimes  \sigma^{\alpha\beta}C$-type tetraquark states are taken into account  cannot work.

The predicted mass  $M_{\bar{s}s,{\rm 1S}}=3.84\pm0.16 \,\rm{GeV}$ overlaps with the experimental value $M_{X(3915)}=3918.4\pm 1.9\,\rm{MeV}$ slightly \cite{PDG}, the $X(3915)$ cannot be a pure $C\sigma_{\alpha\beta} \otimes  \sigma^{\alpha\beta}C$-type $cs\bar{c}\bar{s}$ tetraquark state. The predicted mass  $M_{\bar{s}s,{\rm 2S}}=4.40\pm0.09 \,\rm{GeV}$ overlaps with the experimental value  $ M_{X(4500)} = 4506 \pm 11 ^{+12}_{-15} \mbox{ MeV}$  slightly \cite{LHCb-4500-1606}, the $X(4500)$ cannot be a pure $C\sigma_{\alpha\beta} \otimes  \sigma^{\alpha\beta}C$-type $cs\bar{c}\bar{s}$ tetraquark state. As the central values of the  $M_{\bar{s}s,{\rm 1S}}$ and $M_{\bar{s}s,{\rm 2S}}$ differ from the central values of the $ M_{X(3915)}$ and $ M_{X(4500)}$ significantly, it is difficult to assign   the $ M_{X(3915)}$ and $ M_{X(4500)}$ to be the $C\sigma_{\alpha\beta} \otimes  \sigma^{\alpha\beta}C$-type $cs\bar{c}\bar{s}$ tetraquark states. The $X_{\bar{s}s,{\rm 1S}}$ and $X_{\bar{s}s,{\rm 2S}}$ are new particles, the present predictions can be confronted to the experimental data in the future.

\section{Conclusion}
In this article, we study the ground states and the first radial excited states of the   $C\sigma_{\alpha\beta} \otimes  \sigma^{\alpha\beta}C$-type hidden-charm tetraquark states with the QCD sum rules  by calculating the contributions of the vacuum condensates up to dimension 10 in a consistent way. We separate the ground state contributions from the first radial excited state contributions unambiguously, and obtain the QCD sum rules for the ground states and the first radial excited states respectively.
Then we  search for the  Borel parameters and continuum threshold
parameters    according to  the  four criteria: $\bf{1_\cdot}$ Pole dominance at the hadron side;
$\bf{2_\cdot}$ Convergence of the operator product expansion;
$\bf{3_\cdot}$ Appearance of the Borel platforms;
$\bf{4_\cdot}$ Satisfying the energy scale formula. Finally, we obtain the masses and pole residues of the $C\sigma_{\alpha\beta} \otimes  \sigma^{\alpha\beta}C$-type hidden-charm tetraquark states. The masses can be confronted to the experimental data in the future, while the pole residues can be used to study the relevant processes  with the three-point QCD sum rules or the light-cone QCD sum rules.

\section*{Acknowledgements}
This  work is supported by National Natural Science Foundation,
Grant Numbers 11375063,  and Natural Science Foundation of Hebei province, Grant Number A2014502017.

\end{document}